%% file: ms.tex
\documentclass[master]{ws-rv9x6} % 'master' to combine all chapters
\usepackage{ws-rv-har}    % author-date citation/references %%% WS 04-Dec-17
\usepackage{ws-rv-thm}    % comment this line when `amsthm / theorem / ntheorem` package is used
\usepackage{subfigure}    % required only when side-by-side / subfigures are used
\usepackage{ws-index}     % required only when multiple indexes are used
\usepackage{enumitem}
\usepackage[percent]{overpic}
\usepackage{xcolor}
\makeindex
\newindex{aindx}{adx}{and}{Author Index}       % author index
\renewindex{default}{idx}{ind}{Subject Index}  % subject index

\begin{document}		
\newcommand{\ltsima}{$\; \buildrel < \over \sim \;$}
\newcommand{\lsim}{\lower.5ex\hbox{\ltsima}}
\newcommand{\gtsima}{$\; \buildrel > \over \sim \;$}
\newcommand{\gsim}{\lower.5ex\hbox{\gtsima}}
\newcommand{\bra}{\langle}
\newcommand{\ket}{\rangle}
\newcommand{\lprime}{\ell^\prime}
\newcommand{\lpp}{\ell^{\prime\prime}}
\newcommand{\mprime}{m^\prime}
\newcommand{\mpp}{m^{\prime\prime}}
\newcommand{\ci}{\mathrm{i}}
\newcommand{\dd}{\mathrm{d}}
\newcommand{\veck}{\mathbf{k}}
\newcommand{\vecx}{\mathbf{x}}
\newcommand{\vecr}{\mathbf{r}}
\newcommand{\vecv}{\mathbf{\upsilon}}
\newcommand{\vecw}{\mathbf{\omega}}
\newcommand{\vecj}{\mathbf{j}}
\newcommand{\vecq}{\mathbf{q}}
\newcommand{\vecl}{\mathbf{l}}
\newcommand{\vecn}{\mathbf{n}}
\newcommand{\lm}{\ell m}
\newcommand{\that}{\hat{\theta}}
\newcommand{\thatp}{\that^\prime}
\newcommand{\chip}{\chi^\prime}
\newcommand{\hs}{\hspace{1mm}}
\newcommand{\nar}{New Astronomy Reviews}
\def\gsim{~\rlap{$>$}{\lower 1.0ex\hbox{$\sim$}}}
\def\lsim{~\rlap{$<$}{\lower 1.0ex\hbox{$\sim$}}}
\def\Msun {\,\mathrm{M}_\odot}
\def\Jcrit {J_\mathrm{crit}}
\newcommand{\rsun}{R_{\odot}}
\newcommand{\mbh}{M_{\rm BH}}
\newcommand{\Msunyr}{M_\odot~{\rm yr}^{-1}}
\newcommand{\mdot}{\dot{M}_*}
\newcommand{\ledd}{L_{\rm Edd}}
\newcommand{\cmc}{{\rm cm}^{-3}}
\def\gsim{~\rlap{$>$}{\lower 1.0ex\hbox{$\sim$}}}
\def\lsim{~\rlap{$<$}{\lower 1.0ex\hbox{$\sim$}}}
\def\Msun {\,\mathrm{M}_\odot}
\def\Jcrit {J_\mathrm{crit}}

\def\simgreat{\lower2pt\hbox{$\buildrel {\scriptstyle >}
   \over {\scriptstyle\sim}$}}
\def\simless{\lower2pt\hbox{$\buildrel {\scriptstyle <}
   \over {\scriptstyle\sim}$}}
\def\msobh{M_\bullet^{\rm sBH}}
\def\zodot{\,{\rm Z}_\odot}
\newcommand{\lambdabar}{\mbox{\makebox[-0.5ex][l]{$\lambda$} \raisebox{0.7ex}[0pt][0pt]{--}}}

\def\na{NewA}%
          % New~Astronomy
\def\aj{AJ}%
          % Astronomical Journal
\def\araa{ARA\&A}%
          % Annual Review of Astron and Astrophys
\def\apj{ApJ}%
          % Astrophysical Journal
\def\apjl{ApJ}%
          % Astrophysical Journal, Letters
\def\jcap{JCAP}

\def\pasa{PASA}

\def\apjs{ApJS}%
          % Astrophysical Journal, Supplement
\def\ao{Appl.~Opt.}%
          % Applied Optics
\def\apss{Ap\&SS}%
          % Astrophysics and Space Science
\def\aap{A\&A}%
          % Astronomy and Astrophysics
\def\aapr{A\&A~Rev.}%
          % Astronomy and Astrophysics Reviews
\def\aaps{A\&AS}%
          % Astronomy and Astrophysics, Supplement
\def\azh{AZh}%
          % Astronomicheskii Zhurnal
\def\baas{BAAS}%
          % Bulletin of the AAS
\def\jrasc{JRASC}%
          % Journal of the RAS of Canada
\def\memras{MmRAS}%
          % Memoirs of the RAS
\def\mnras{MNRAS}%
          % Monthly Notices of the RAS
\def\pra{Phys.~Rev.~A}%
          % Physical Review A: General Physics
\def\prb{Phys.~Rev.~B}%
          % Physical Review B: Solid State
\def\prc{Phys.~Rev.~C}%
          % Physical Review C
\def\prd{Phys.~Rev.~D}%
          % Physical Review D
\def\pre{Phys.~Rev.~E}%
          % Physical Review E
\def\prl{Phys.~Rev.~Lett.}%
\def\pasp{PASP}%
          % Publications of the ASP
\def\pasj{PASJ}%
          % Publications of the ASJ
\def\qjras{QJRAS}%
          % Quarterly Journal of the RAS
\def\skytel{S\&T}%
          % Sky and Telescope
\def\solphys{Sol.~Phys.}%
          % Solar Physics

          % Solar Physics
\def\sovast{Soviet~Ast.}%
          % Soviet Astronomy
\def\ssr{Space~Sci.~Rev.}%
          % Space Science Reviews
\def\zap{ZAp}%
          % Zeitschrift fuer Astrophysik
\def\nat{Nature}%
          % Nature
\def\iaucirc{IAU~Circ.}%
          % IAU Cirulars
\def\aplett{Astrophys.~Lett.}%
          % Astrophysics Letters
\def\apspr{Astrophys.~Space~Phys.~Res.}%
          % Astrophysics Space Physics Research
\def\bain{Bull.~Astron.~Inst.~Netherlands}%
          % Bulletin Astronomical Institute of the Netherlands
\def\fcp{Fund.~Cosmic~Phys.}%
          % Fundamental Cosmic Physics
\def\gca{Geochim.~Cosmochim.~Acta}%
          % Geochimica Cosmochimica Acta
\def\grl{Geophys.~Res.~Lett.}%
          % Geophysics Research Letters
\def\jcp{J.~Chem.~Phys.}%
          % Journal of Chemical Physics
\def\jgr{J.~Geophys.~Res.}%
          % Journal of Geophysics Research
\def\jqsrt{J.~Quant.~Spec.~Radiat.~Transf.}%
          % Journal of Quantitiative Spectroscopy and Radiative Trasfer
\def\memsai{Mem.~Soc.~Astron.~Italiana}%
          % Mem. Societa Astronomica Italiana
\def\nphysa{Nucl.~Phys.~A}%

\def\physrep{Phys.~Rep.}%
          % Physics Reports
\def\physscr{Phys.~Scr}%
          % Physica Scripta
\def\planss{Planet.~Space~Sci.}%
          % Planetary Space Science
\def\procspie{Proc.~SPIE}%
          % Proceedings of the SPIE

\newcommand{\rmp}{Rev. Mod. Phys.}
\newcommand{\ijmpd}{Int. J. Mod. Phys. D}
\newcommand{\sovjetp}{Soviet J. Exp. Theor. Phys.}
\newcommand{\jkas}{J. Korean. Ast. Soc.}
\newcommand{\PPVI}{Protostars and Planets VI}
\newcommand{\njp}{New J. Phys.}
\newcommand{\rap}{Res. Astro. Astrophys.}

\input{ch8.tex}

{
\bibliographystyle{ws-rv-har}    % author-date citation/references %%% WS 04-Dec-17
\bibliography{ref}
}

\printindex[aindx]           % to print author index
\printindex                  % to print subject index

\end{document}

%% file: ch8.tex
%%%%%%%%%%%%%%%%%%%%%%%%%%%%%%%%%%%%%%%%%%%%%%%%%%%%%%%%%%%%%%%%%%%%%%%%%%
%% Review Volume (last updated on 20-4-2015)                            %%
%% Trim Size: 9in x 6in                                                 %%
%% Text Area: 7.35in (include runningheads) x 4.5in                     %%
%% Main Text: 10 on 13pt                                                %%
%% For support: Yolande Koh, <ykoh@wspc.com.sg>                         %%
%%              D. Rajesh Babu, <rajesh@wspc.com.sg>                    %%
%%%%%%%%%%%%%%%%%%%%%%%%%%%%%%%%%%%%%%%%%%%%%%%%%%%%%%%%%%%%%%%%%%%%%%%%%%
%%
%\documentclass[wsdraft]{ws-rv9x6} % to draw border line around text area
%\documentclass{ws-rv9x6}
%\usepackage{subfigure}   % required only when side-by-side / subfigures are used
%\usepackage{ws-rv-thm}   % comment this line when `amsthm / theorem / ntheorem` package is used
%\usepackage{ws-rv-van}   % numbered citation & references (default)
%%\usepackage{ws-index}   % to produce multiple indexes
%\makeindex
%\newindex{aindx}{adx}{and}{Author Index}       % author index
%\renewindex{default}{idx}{ind}{Subject Index}  % subject index

%\newcommand{\Msun}{M_{\odot}}
%

%\begin{document}

\title{Formation of the First Black Holes}

\setcounter{chapter}{7}
\chapter[Primordial Supermassive Stars]{Formation of the First Black Holes: \\ Evolution and Final Fates of Rapidly Accreting Supermassive Stars$^1$}
\label{super}

\author[Takashi Hosokawa]{Takashi Hosokawa}
%\index[aindx]{Author, F.} % or \aindx{Author, F.}
%\index[aindx]{Author, S.} % or \aindx{Author, S.}

\address{Department of Physics, Kyoto University, Kyoto, 606-8502, Japan \\
hosokawa@tap.scphys.kyoto-u.ac.jp}

\begin{abstract}
 The formation of supermassive stars (SMSs) is a possible pathway to seed supermassive black holes in the early universe. This chapter summarizes recent theoretical efforts to understand their evolution, highlighting effects of very rapid accretion at the rates of $\dot{M}_* \gtrsim 0.1~$M$_\odot$~yr$^{-1}$. Stellar evolution calculations predict that such an accreting SMS has a characteristic feature, i.e., a very large radius that monotonically increases as the stellar mass increases. The radius exceeds $7000~\rsun$ (or $30$~AU) after the star accretes the gas of more than $10^3$~M$_\odot$. We show that the emergence of the ``supergiant protostar'' stage is a key for the formation of SMSs, because resulting radiative feedback against the accretion flow is substantially weakened during this stage. We also show that these SMSs end their lives while the accretion continues, and that their final fates vary with different accretion rates. 
\end{abstract}
%\markright{Customized Running Head for Odd Page} % default is Chapter Title.
\body
\setcounter{page}{145}

\footnotetext{$^1$ Preprint~of~a~review volume chapter to be published in Latif, M., \& Schleicher, D.R.G., ''Evolution and Final Fates of Rapidly Accreting Supermassive Stars'', Formation of the First Black Holes, 2018 \textcopyright Copyright World Scientific Publishing Company, https://www.worldscientific.com/worldscibooks/10.1142/10652 }

%\tableofcontents

%--------------------------------------%
% maximum length: 20 pages in total
%
%-------------------------------------------------------------------------------%
%1 Introduction to black holes and high-redshift structure formation
%1.1 Astrophysical Black Holes
%1.2 Cosmological structure formation
%1.3 Thermodynamics and chemistry of the early Universe
%1.4 The formation of the first cosmological objects
%1.5 The dynamics of stellar clusters
%1.6 Supermassive stars and quasi-stars as black hole progenitors
%
%2 Black hole formation scenarios
%2.1 Formation of the first stars as progenitors of the first black holes
%2.2 Black hole formation in the first stellar clusters
%2.3 Black hole formation via gas-dynamical processes (direct collapse)
%2.4 Statistical predictions for the first black holes
%2.5 Growth and feedback from the first black holes
%
%3 Observations and future prospects
%3.1 Current status concerning high-redshift black holes
%3.2 Low-mass black holes in dwarf galaxies as probes of the formation mechanism
%3.3 Future probes via gravitational waves
%3.4 Prospects for ATHENA, JWST and LSST
%-----------------------------------------------------------------------------------%

%%%%%%%%%%%%%%%%%%%%%%%%%%%%%%%%%%%%%%%%%%%%%%%%%%%%%%%%%%%%%
\section{Introduction}
\label{ra_sec1}
%%%%%%%%%%%%%%%%%%%%%%%%%%%%%%%%%%%%%%%%%%%%%%%%%%%%%%%%%%%%%

The formation and evolution of $\gtrsim 10^5~\Msun$ supermassive stars (SMSs) has been intensively studied in recent years.  It has been predicted that such SMSs will leave equally massive black holes (BHs) after their deaths \citep{Iben1963, Chandra1964}. The massive BHs are favored as massive seeds to explain the quick formation of the SMBHs, which has been suggested by recent discoveries of  $> 10^9~ \Msun$ supermassive black holes (SMBHs)  at high red-shifts $z \gtrsim 6$.

%-----------------------------------------------------------------------%
% A great motivation of such studies is to elucidate the origin of $\gtrsim 10^9~\Msun$ supermassive black holes (SMBHs) at high redshifts $z \gtrsim 6$.

Nonetheless, it is still uncertain whether the formation of the SMSs surely occurs in the early universe. It is thought that the typical mass of the normal primordial stars is $\sim 100~\Msun$\citep{Hirano2015,Hirano2014,Susa14,Hosokawa11}, see also discussion in chapter~4. Hence various authors are exploring plausible processes to cause the formation of SMSs, if any, which only rarely occurs under some non-standard conditions.  
One of them is the so-called direct collapse (DC) scenario \citep{Bromm03}, for which a number of studies have been conducted in the past decade, and which was described here in chapter~5, while additional pathways were outlined in chapter~7. We here focus mostly on one key process in the DC model, the stellar mass growth via rapid accretion.
In the star formation process in general, including the DC model, a tiny protostar first appears after the gravitational collapse of a gas cloud \citep{Omukai2000,Omukai2001,Inayoshi2014}.
Such an embryonic protostar then grows in mass via gas accretion from a surrounding envelope. 
For the SMS formation through the protostellar accretion, very high rates 
of $\gtrsim 0.1~\Msunyr$ are required because the duration of the evolution 
is limited by the stellar lifetime, $\sim$~Myr.
The DC model provides such extremely rapid accretion, and it is considered as a possible process to yield the SMSs.

\begin{table}[htp]
\caption{default}
\begin{center}
\begin{tabular}{|c|c|}
Abbreviation & full name\\ \hline
SMS & supermassive star\\
SMBH & supermassive black hole\\
BH & black hole\\
DC & direct collapse\\
KH & Kelvin-Helmholtz\\
ZAMS & zero-age-main-sequence\\
GR & general relativistic
\end{tabular}
\end{center}
\label{abrsuper}
\end{table}%

%----------------------------------------------------------------------------%

The evolution and final fates of the SMSs had been investigated by various authors  \citep{Osaki66,Unno71,Fricke73,Fuller86}, but they did not consider the realistic formation process, i.e., the rapid gas accretion to obtain its huge mass. Normally a SMS in the stable nuclear burning phase was just assumed as an initial condition, and its subsequent evolution was followed until its death. Considering the formation process with the rapid accretion, however, the timescale of the stellar mass growth is actually comparable to the stellar lifetime. Therefore, the birth and death of the SMSs cannot be separately investigated.

%----------------------------------------------------------------------------%

The stellar radiative feedback against the accretion flow is a potential obstacle for the formation of massive stars. In the normal primordial star formation, where the accretion rates are $\sim 10^{-2}-10^{-3}~\Msunyr$, the stellar emissivity of the H-ionizing radiation rapidly rises 
as the star approaches the zero-age main-sequence (ZAMS) stage.
The resulting UV feedback caused by the formation and expansion of an HII region easily halts the accretion toward the star \citep{Mckee2008,Hosokawa11}. 
Although with some variations with different accretion rates, this occurs before the stellar mass reaches $10^3~\Msun$ in many cases. 
In the DC model, which postulates the SMS formation, the stellar bolometric luminosity should
be much higher than in the normal primordial cases, 
exceeding $10^8~L_{\odot}$ for $M_* \gtrsim 10^4~\Msun$ stars. 
If the SMS emits a copious amount of ionizing photons, tremendous UV radiative feedback may shut off the accretion way before the formation of the SMSs.
Solving the evolution of a rapidly accreting SMS is necessary to evaluate the strength of the UV feedback.

%-----------------------------------------------------------------------------%

In the same vein, the final fate of the SMS, i.e., at which point the star collapses into a BH, is also affected by the mass accretion. The BH left after the death of the SMS sets the starting point for the further mass growth toward a SMBH. This is of great importance to elucidate the origin of the SMBHs.

%%%%%%%%%%%%%%%%%%%%%%%%%%%%%%%%%%%%%%%%%%%%%%
\section{Modeling of Accreting Protostars}
\label{ra_sec2}
%%%%%%%%%%%%%%%%%%%%%%%%%%%%%%%%%%%%%%%%%%%%%%

Numerically modeling the evolution of accreting stars was initiated in studies on the binary evolution which causes rapid mass exchange among the stars \citep{Kippenhahn77,Neo77}.
The basic numerical technique was then extended and applied for studies on the star formation by Steven Stahler and his collaborators \citep{SST80,Stahler88}.
They performed a series of pioneering calculations for the present-day low-mass ($\sim 1~\Msun$) star formation, considering simple accretion histories with constant rates of $\mdot \sim 10^{-6 \sim -5}~\Msunyr$.
Then basically the same method was extensively applied for the evolution of more massive 
protostars growing with the higher accretion rates \citep{Palla91,Behrend01,Hosokawa2009,Haemmerle16}. 
In parallel, the evolution of the primordial protostars was also investigated \citep{SPS86,OmukaiPalla2003}. The high accretion rates of $\mdot \sim 10^{-4 \sim -2}~\Msunyr$ were commonly used in these studies on the present-day and primordial star formation. 
Effects of the extremely rapid mass accretion with $\mdot \gtrsim 10^{-2}~\Msunyr$ have been examined recently, motivated by the developments of the DC model.

%------------------------------------------------------------------------------%

Regardless of the great differences of the considered accretion rates, the above studies deal with the same governing equations to construct stellar models, 
\begin{equation}
\left( \frac{\partial r}{\partial M} \right)_t = \frac{1}{4 \pi \rho r^2},
\label{eq:con} 
\end{equation}
\begin{equation}
\left( \frac{\partial P}{\partial M}  \right)_t = - \frac{GM}{4 \pi r^4}, 
\label{eq:mom}
\end{equation}
\begin{equation}
\left( \frac{\partial L}{\partial M} \right)_t 
= \epsilon - T \left( \frac{\partial s}{\partial t}  \right)_M ,
\label{eq:ene}
\end{equation}
\begin{equation}
\left( \frac{\partial s}{\partial M} \right)_t
= \frac{G M}{4 \pi r^4} \left( \frac{\partial s}{\partial p} \right)_T
  \left( \frac{L}{L_s} - 1  \right) C ,
\label{eq:heat}
\end{equation}                                                                
which are essentially the four stellar structure equations taking into account effects of the mass accretion.
In the above, $M$ is the mass coordinate, $\epsilon$ the energy
production rate via nuclear fusion, $s$ the specific entropy,
$L_s$ the radiative luminosity with adiabatic temperature
gradient, $C$ the coefficient determined by the mixing-length
theory when $L > L_s$ ($C=1$ otherwise), and 
the remaining quantities have ordinary meanings.

%-----------------------------------------------------------------%

Effects of the mass accretion are incorporated in the term
$T (\partial s / \partial t)_M$ in Eq.(\ref{eq:ene}), which
is approximately discretized as
\begin{equation} 
\left( \frac{\partial s}{\partial t}  \right)_M = 
  \left( \frac{\partial s}{\partial t} \right)_m 
+ \dot{M}_\ast  \left( \frac{\partial s}{\partial m}  \right)_t ,
\label{eq:relm}
\end{equation}
where $m$ is the inverse mass coordinate measured from the surface, $m \equiv M_{\ast} - M$.
An evolutionary calculation starts with an arbitrary initial
model with small mass, and advances with updating models with
the mass increment via accretion $\mdot \Delta t$, where 
$\Delta t$ is the time step. 

%-------------------------------------------------------------%

This sort of modeling has an uncertainty in how to determine the thermal state of the accreting gas. A self-consistent treatment is possible only when assuming the spherical symmetry not only for the star but also for the surrounding accretion flow. In this case, 
the accretion flow directly hits the stellar surface forming an accretion shock. 
The jump conditions at the shock front consistently determine the thermal state of the gas which settles down into the star in a post-shock layer \citep{SST80}. However, the accreting gas should have some angular momentum, and it will be more realistic to consider the gas accumulating onto the star through an accretion disk. 
A practical way of modeling is to solve the 1D structure of the central star employing the normal photospheric outer boundary conditions, supposing that most of the stellar surface freely radiates while the accreting gas approaches the star through a geometrically thin disk. 
A simple ansatz often used for this case is the so-called ``cold accretion'' limit, where the accreting gas is assumed to have the same specific entropy as in the stellar atmosphere.  
Remember that this is an extreme limit; the cold accretion will be a poor approximation especially with the rapid accretion. In reality, the accreting gas should be somewhat ``warmer'' because it carries a part of the liberated gravitational energy (i.e., the accretion energy) into the star before radiating away. 
For modeling this effect, we temporarily deposit the energy $L_{\rm *,acc}$ into the stellar interior, 
\begin{equation}
L_{\rm *, acc} = \eta L_{\rm acc} \equiv \eta \left( \frac{G M_* \dot{M}_*}{R_*} \right)^{-1},
\end{equation}
where $\eta$ is a free-parameter satisfying $0 \leq \eta \leq 1$.
The above mentioned spherical accretion corresponds to a case with non-zero $\eta$ (but $\eta \neq 1$), whose value is given by solving the flow structure across the shock.
In general, even a small value of $\eta$ (e.g., $\eta \sim 0.01$) is enough to change the evolution from the cold accretion limit. Fortunately, in our cases considered below, uncertainties with varying $\eta$ only appear in an early stage of the evolution (see Section~\ref{ra_sec3}).

%%%%%%%%%%%%%%%%%%%%%%%%%%%%%%%%%%%%%%%%%%%%%%%%%%%%%%%%%%%%
\section{Evolution of Rapidly Accreting Supermassive Stars}
\label{ra_sec3}
%%%%%%%%%%%%%%%%%%%%%%%%%%%%%%%%%%%%%%%%%%%%%%%%%%%%%%%%%%%%

\subsection{Emergence of the ``Supergiant Protostar'' Stage}

\begin{figure}
\centerline{\includegraphics[width=9.5cm]{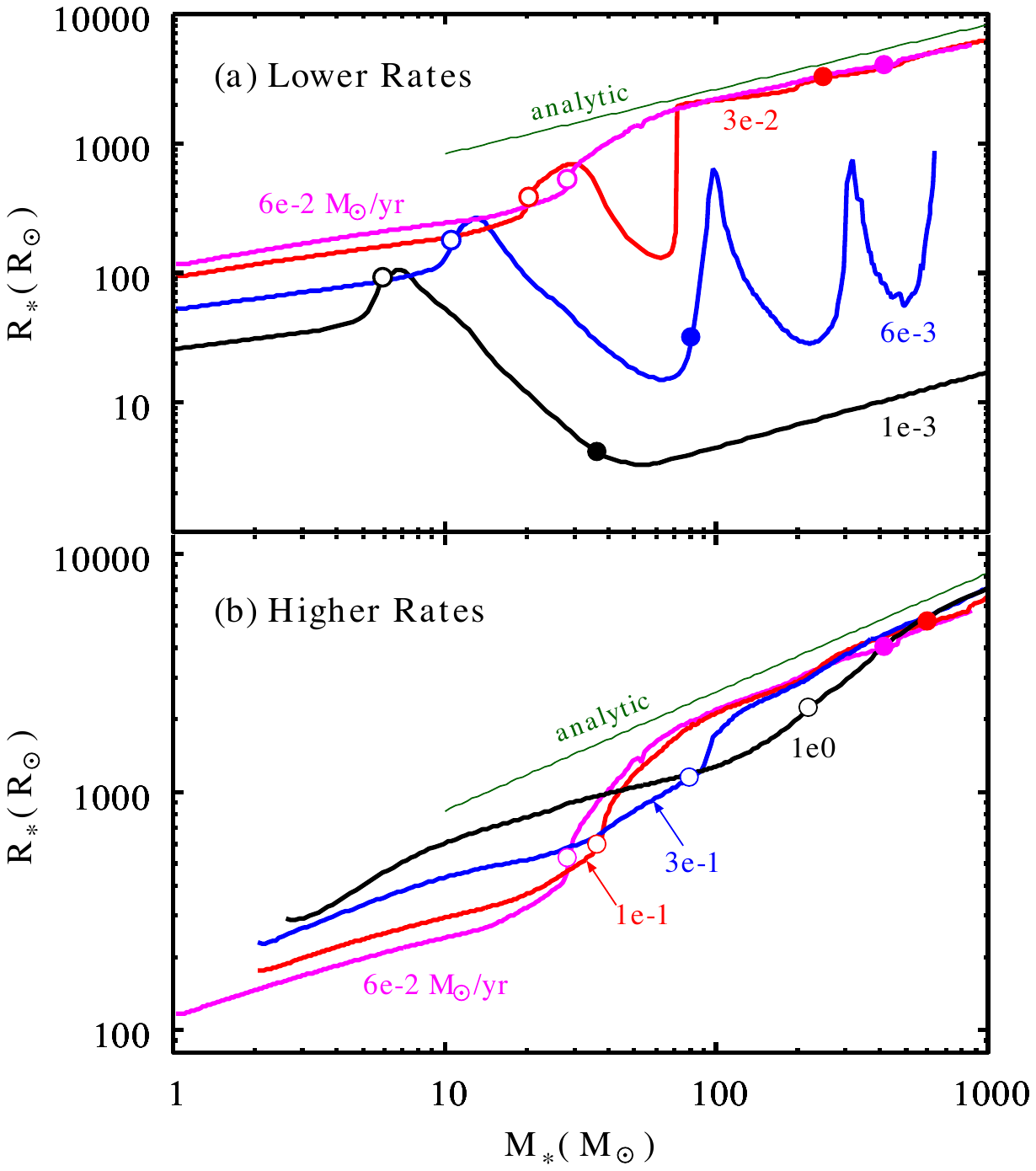}}
\caption{Evolution of the stellar radius with the different accretion rates \citep{Hosokawa12}. The upper and lower panels show cases with the rates lower and higher than $6 \times 10^{-2}~\Msunyr$. The accretion rate represented by each line is labeled in the figure. The thin green line represents the analytic mass-radius relationship given by equation (\ref{ra_anal}). The filled circle on each line represents the epoch when the hydrogen burning starts at the center. Figure adopted from \citet{Hosokawa12},  \textcopyright AAS. Reproduced with permission.} 
\label{ra_fig1}
\end{figure}

We from here concentrate on the protostellar evolution with accretion rates $\gtrsim 10^{-2}~\Msunyr$, through which the formation of the SMSs is supposed to occur. Figure~\ref{ra_fig1} summarizes the results by \citet{Hosokawa12}, who have performed the stellar evolution calculations assuming the spherical accretion (see Section~\ref{ra_sec2}).
The black curve in panel (a) represents the case with $\mdot = 10^{-3}~\Msunyr$, a relatively low accretion rate which expected for the normal primordial star formation. It is helpful for a later reference to describe this case first. A striking feature here is the contraction stage which occurs for the mass range of $6~\Msun \lesssim M_* \lesssim 30~\Msun$.
It corresponds to the so-called Kelvin-Helmholtz (KH) contraction, during which the stellar interior temperature rises as the star loses its energy by radiating away. 
The hydrogen burning starts near the center just after the stellar mass exceeds $30~\Msun$. After that point, the stellar radius turns to increase almost following the mass-radius relation for zero-age main-sequence (ZAMS) stars.
The stellar effective temperature and emissivity of ionizing photons substantially increases during the KH contraction stage. 
The UV feedback against the accretion flows begins to operate from a late stage of such a contraction. The presence of the KH contraction toward the ZAMS stage thus marks the onset the stellar UV feedback. 

%-----------------------------------------------------------------%

Meanwhile, the above evolution drastically changes with the more rapid accretion. Panel (a) shows that, with the higher accretion rate, the KH contraction stage only spans the narrower mass range. 
The contraction turns to the expansion at some point for the accretion rates $6 \times 10^{-3}~\Msunyr$ and $3 \times 10^{-2}~\Msunyr$.
With the highest rate $6 \times 10^{-2}~\Msunyr$, the contraction stage never appears and the stellar radius monotonically increases with increasing the stellar mass.
The stellar radius finally reaches $\simeq 7000~\rsun$ (or nearly 30~AU) when the star has accreted the gas of $1000~\Msun$ in this case. 
Panel (b) summarizes the evolution with the even higher accretion rates, $6 \times 10^{-2}~\Msunyr \leq \mdot \leq 1~\Msunyr$, showing further characteristic features; the radius monotonically increases with the mass for all the cases, and moreover, the lines representing the different accretion rates converge to the same mass-radius relation. 
The radius at $M_* \simeq 1000~\Msun$ is consequently $\simeq$ 30~AU for all the cases, regardless of the different accretion rates.

%-----------------------------------------------------------------%

The mass-radius relation toward which the lines finally converge is approximately given by
\begin{equation}
R_* \simeq 2.6 \times 10^3~R_\odot 
\left( \frac{M_*}{100~M_\odot} \right)^{1/2} ,
\label{ra_anal}
\end{equation}
which is analytically derived as follows. We begin with the general expression of the stellar luminosity
\begin{equation}
 L_* = 4 \pi R_*^2 \sigma T_{\rm eff}^4,
\label{ra_eqlum}
\end{equation}
where $\sigma$ is the Stefan-Boltzmann constant.
As for very massive stars, the luminosity is very close to the Eddington value,
\begin{equation}
 L_* \simeq L_{\rm Edd} \propto M_* .
\label{ra_eqledd}
\end{equation}
In the current cases with the high accretion rates, a stellar surface layer is difficult to contract because it always traps a part of the heat released from the contracting interior. The surface layer consequently has a high specific entropy to greatly inflate. The effective temperature accordingly drops to several $\times$ 1000~K, where H$^-$ bound-free absorption becomes the dominant process to provide the opacity. 
Since the H$^-$ opacity has a very strong temperature-dependence, the effective temperature is almost fixed at a constant value. Substituting equation (\ref{ra_eqledd}) and $T_{\rm eff} \simeq 5000$~K into equation (\ref{ra_eqlum}) provides equation (\ref{ra_anal}).
The above derivation includes no dependence of the accretion rate, which explains why the different curves in Figure~\ref{ra_fig1} finally converge into the same mass-radius relation.

%------------------------------------------------------------------------%

\subsection{Evolution toward the Supermassive Stars}

\begin{figure}
\centerline{\includegraphics[width=9.5cm]{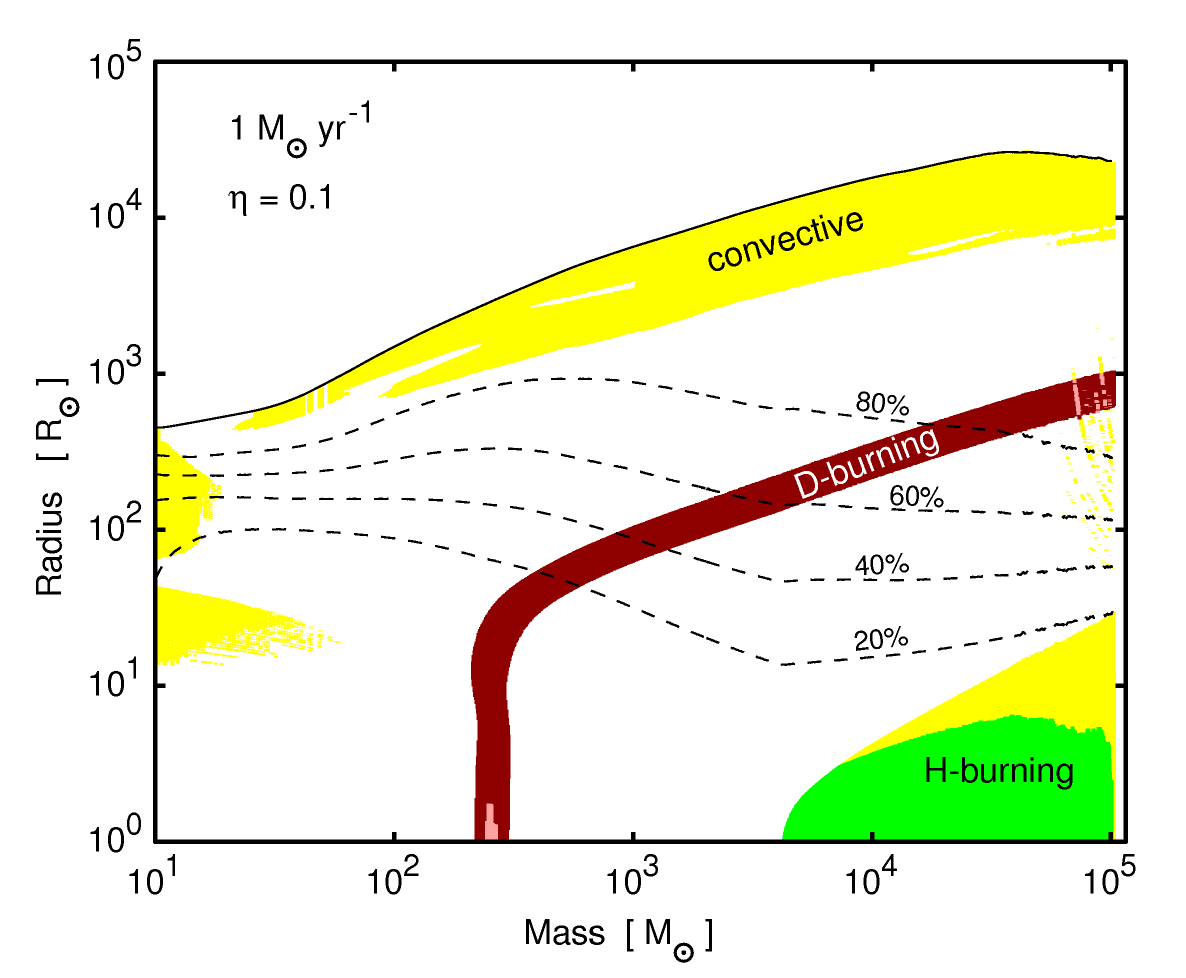}}
\caption{Evolution of the stellar interior structure 
until the accreted stellar mass reaches $10^5~\Msun$
for the accretion rate $\mdot = 1~\Msunyr$ and 
energy fraction $\eta = 0.1$ \citep{Hosokawa2013}.
The black solid and dashed lines represent the radial positions 
of the stellar surface and mass coordinates of $80\%$, $60\%$, $40\%$, and $20\%$ of the total
stellar mass in descending order. The white and yellow areas represent radiative and convective 
zones without nuclear fusion, respectively. 
The brown stripe denotes the radiative layer where deuterium burning occurs.
The green area represents a convective core where hydrogen 
burning occurs. The pink zones depict convective deuterium burning. Figure adopted from \citet{Hosokawa2013},  \textcopyright AAS. Reproduced with permission.} 
\label{ra_fig2}
\end{figure}

\begin{figure}
\centerline{\includegraphics[width=11.5cm]{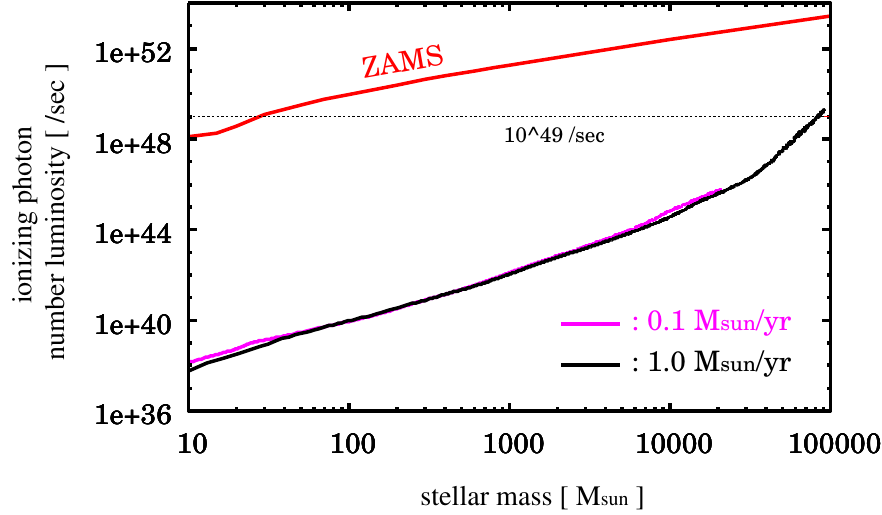}}
\caption{Evolution of the ionizing photon number luminosity of accreting supermassive stars. The black and magenta lines represent the cases with different constant accretion rates of $1~\Msunyr$ and $0.1~\Msunyr$. The red line represents the reference case of non-accreting zero-age main-sequence stars.} 
\label{ra_fig3}
\end{figure}

\citet{Hosokawa2013} have further extended calculations to follow the later evolution until the stellar mass reaches $10^4 \sim 10^5~\Msun$.  
As the outer boundary of the stellar models, the disk accretion described in Section~\ref{ra_sec2} is adopted. The parameter $\eta$ is set to be 0.1, i.e., 10\% of the accretion energy is assumed to be deposited into the stellar interior. 
Figure~\ref{ra_fig2} illustrates the evolution of the stellar interior structure with the constant accretion rate of $1~\Msunyr$. We confirm that equation (\ref{ra_anal}) still describes the evolution of the radius well for $100~\Msun \lesssim M_* \lesssim 1000~\Msun$, in spite of the different prescriptions of the spherical and disk accretion.
Although not presented here, effects of varying the parameter $\eta$ only appear in an early evolution for $M_* \lesssim 100~\Msun$. The same evolution through the supergiant protostar appears even with $\eta = 0$, the extreme case of the cold accretion. This has been also confirmed by other authors using different numerical codes~\citep{Umeda2016,Haemmerle17}. 

%------------------------------------------------------------------------%

According to Figure~\ref{ra_fig2}, the monotonic increase of the stellar radius continues after the star accretes $1000~\Msun$. The analytic formula given by equation (\ref{ra_anal}) still provides a good fit for the evolution until the stellar mass reaches $\sim 10^4~\Msun$. 
The increase of the radius slows down after that, and finally saturates at $M_* \simeq 3 \times 10^4~\Msun$. The star slightly contracts as the mass increases for $3 \times 10^4~\Msun \lesssim M_* \lesssim 10^5~\Msun$. 
The deviation from equation (\ref{ra_anal}) is caused by the decrease of the density in the stellar atmosphere, with which the H$^-$ ions cease to dominate the opacity. The similar deviation has been also reported using the different codes~\citep{Haemmerle17}. 

%------------------------------------------------------------------------%

Figure~\ref{ra_fig2} shows that a surface layer which only has a small fraction of the total mass significantly inflates to cover a large part of the stellar radial extent. The star has a bloated envelope and compact core like red giants. Most of the stellar mass is contained in the central core, which rather contracts as the stellar mass increases. The central temperature gradually rises as in the ordinary KH contraction. As a result, the nuclear burning initiates when the total stellar mass reaches $\simeq 4000~\Msun$. The central convective core gradually extends after that, because of the heat input by the nuclear fusion. The evolution of the radius does not follow the mass-radius relation for the ZAMS stars. The outer convective layer remains bloated as long as the rapid mass accretion continues. 
Such a characteristic structure of the rapidly accreting protostars (e.g., very large radius) has been also predicted by a naive analytic argument \citep{Schleicher13}. However, recall that it is the stellar atmosphere where H$^-$ absorption dominates the opacity that regulates the evolution of the radius. 
Thus numerical modeling resolving the detailed structure near the surface is indispensable to accurately describe the stellar interior structure and resulting evolution.

%-------------------------------------------------------------------------%

As described above, SMSs with extremely rapid mass accretion inflate to have the cool atmosphere. The resulting effective temperature is only $\simeq 5000$~K, with which a star only emits a small amount of ionizing photons even after it accretes the gas of $\gg 1000~\Msun$. Figure~\ref{ra_fig3} shows that the UV emissivity of the accreting SMSs is much smaller than that of non-accreting counterparts by several orders of magnitude. 
The UV emissivity gradually increases as the stellar mass increases, because the bolometric luminosity rises roughly in proportion to the mass. Nonetheless, the UV emissivity of a bloating $\sim 10^5~\Msun$ SMS is still comparable to that of a non-accreting star with a few $\times~10~\Msun$.
The resulting UV feedback is too weak to disturb the massive accretion flow toward the accreting SMSs. The formation of SMSs via accretion is thus possible as far as such high accretion rates exceeding $\sim 10^{-2}~\Msunyr$ are maintained. 

%-------------------------------------------------------------------------%

In the calculation described above, the general relativistic (GR) effects are incorporated using the post-Newton approximation, as an effective modification of the gravity constant. However, the GR effects are not strong enough yet to affect the stellar evolution. We describe the final evolutionary stage where the star dies and dynamically collapses in the next section.

%%%%%%%%%%%%%%%%%%%%%%%%%%%%%%%%%%%%%%%%%%%%%%%%%%%%%%%
\section{Final Fates of Accreting Supermassive Stars}
\label{ra_sec4}
%%%%%%%%%%%%%%%%%%%%%%%%%%%%%%%%%%%%%%%%%%%%%%%%%%%%%%%

\begin{figure}
\centerline{\includegraphics[width=9cm,angle=-90]{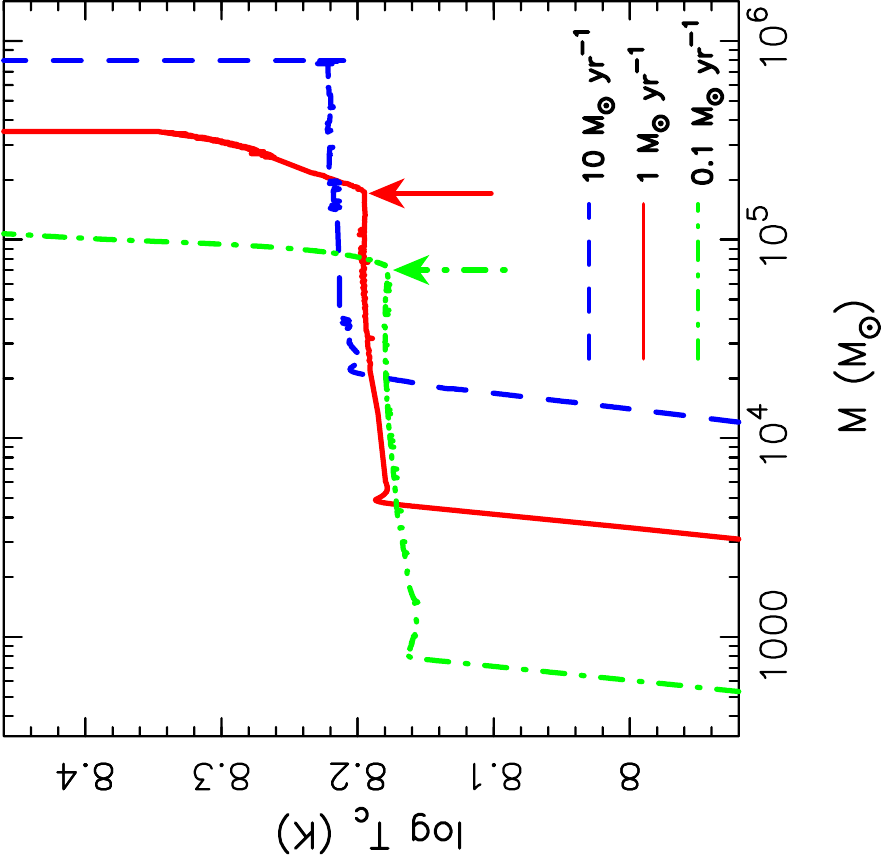}}
\caption{The evolution of the central temperature $T_{\rm c}$ 
as the stellar mass increases \citep{Umeda2016}. 
Only the values around the stable nuclear burning are shown for clarity.
The different lines represent the different accretion rates,
$10~\Msunyr$ (blue dashed), $1~\Msunyr$ (red solid),
and $\dot M = 0.1~\Msunyr$ (green dot-dashed line).
The vertical arrows indicate the points where the central hydrogen
is exhausted with $\dot{M} = 0.1$ and $1~\Msunyr$. Figure adopted from \citet{Umeda2016},  \textcopyright AAS. Reproduced with permission. } 
\label{ra_fig4}
\end{figure}

Figure~\ref{ra_fig4} presents the evolution of the stellar central temperature in the final stage up to the onset of the collapse \citep{Umeda2016}.
The governing equations here are the same as equations (\ref{eq:con}) -- (\ref{eq:heat}), except that equation (\ref{eq:mom}) is modified to consider the post-Newton approximation, and to include the acceleration term to follow the dynamical collapse. Figure~\ref{ra_fig4} shows that the evolution varies with the different accretion rates. However, all the cases commonly present the characteristic stage where the central temperature is almost constant at $\log T_{\rm c} \simeq 8.2$. The hydrogen burning takes place during this period. The epoch of igniting hydrogen depends on the accretion rate, i.e., at the higher stellar mass with the higher accretion rate. 
After the stage of the stable hydrogen burning, the temperature sharply rises for all the cases. 
Although the calculations follow the evolution until the central temperature reaches $\log T_{\rm c} = 9.2$, the evolution after that is too rapid for the stellar mass to increase via the accretion. We regard this epoch as the point where the final mass and fate of the star are determined. 

%-----------------------------------------------------------%

Figure~\ref{ra_fig4} shows that the final mass is higher with the higher accretion rate. Furthermore, the physical process which causes the collapse also differ with the rates. With the lowest rate $0.1~\Msunyr$, the collapse occurs at $M_* \simeq 10^5~\Msun$, until which the mass growth has continued for $\sim$ Myr starting with an embryonic protostar. Its duration is almost comparable to the stellar lifetime. In fact, the helium mass fraction $Y$ is almost zero near the center at the epoch of $\log T_{\rm c} = 8.7$, meaning that the collapse occurs when the nuclear fuel has been exhausted in this case.  
With the higher rate $1~\Msunyr$, on the other hand, the helium fraction $Y$ is still the unity at the same epoch. The collapse is caused by the GR instability at the beginning of the helium burning stage. With the extremely high rate of $10~\Msunyr$, the collapse with the hydrogen mass fraction $X \simeq 0.5$ and $Y=1$.
The GR instability does not cause the collapse until the stellar mass reaches $M_* \simeq 8 \times 10^5~\Msun$ in the course of the hydrogen-burning stage.  

%------------------------------------------------------------%

Since the upper mass for non-accreting SMSs is $\simeq 1.4 \times 10^5~\Msun$ for the initial primordial abundance \citep{ST85}, the above mentioned variations of the final mass are caused by the rapid mass accretion. The interior structure of an accreting SMS greatly differs from that of the non-accreting counterpart with the same mass. In fact, non-accreting SMSs are normally modeled as $n=3$ polytropic star supposing the fully-convective structure. As shown in Figure~\ref{ra_fig2}, however, the rapidly accreting star has a large radiative layer in its interior. 
The numerical results show that the more rapid mass accretion makes the stellar mass distribution less centrally condensed at a given stellar mass \citep{Hosokawa12}. 
This explains why the GR instability triggers the stellar collapse at the higher stellar mass with the more rapid accretion. 
We conclude that the rapid mass accretion influences not only the evolution during the stellar mass growth, but also the final fate when it dies.

%---------------------------------------------------------------------------%

Similar variations of the final mass of accreting SMSs have been recently studied by other authors using different numerical codes \citep{Woods2017,Haemmerle17}. Regardless of some technical differences expected, they all show the similar overall trends; the final mass roughly logarithmically increases with increasing the accretion rate. The quantitative differences by factors of a few are to be examined in future studies.

%%%%%%%%%%%%%%%%%%%%%%%%%%%%%%%%%%%%%%%%%%%%%%%%%%%
\section{Roles of the Supergiant Protostars}
\label{ra_sec4}
%%%%%%%%%%%%%%%%%%%%%%%%%%%%%%%%%%%%%%%%%%%%%%%%%%%

\begin{figure}
\centerline{\includegraphics[width=12cm,angle=0]{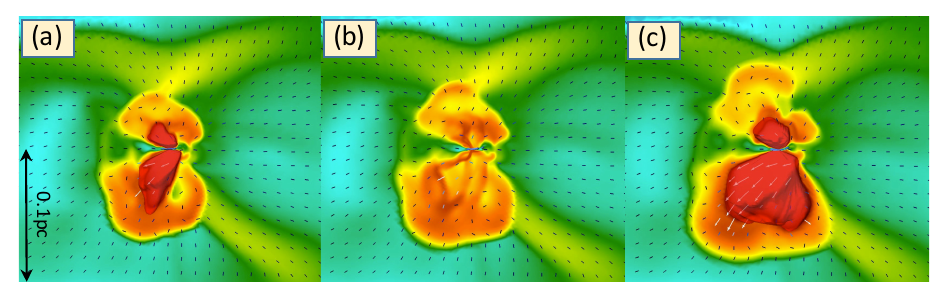}}
\caption{The intermittent UV feedback observed in a 3D radiation hydrodynamic simulation
(case C-HR2-m0 in \citet{Hosokawa2016}). The panels (a)-(c) show a time sequence, where the color scale represents the gas temperature distribution around the central protostar.
Note that an HII region (red) temporarily disappears in panel (b). Figure adopted from \citet{Hosokawa17}, copyright of Societ\`a Astronomica Italiana, reproduced with permission.} 
\label{ra_fig5}
\end{figure}

The stellar evolution calculations described above have adopted simple accretion histories with constant rates for simplicity. In realistic situations, however, the accretion histories are not steady, but variable in time. The mass accretion onto a star should occur through a massive circumstellar disk, where the mass and angular momentum transfer is caused by the gravitational torque \citep{Greif12}. The disk undergoes the gravitational fragmentation in extreme cases. The fragments quickly migrate inward via the gravitational torque to eventually fall onto a star \citep{Inayoshi2014b,Sakurai2016}. It results in an accretion burst event, during which the rate is substantially enhanced. Such a burst is normally followed by a relatively long quiescent phase, where the rate is much lower than the mean value. 

%-------------------------------------------------------------------------------%

The effects of such variable accretion on the evolution of the SMSs have been also studied using the numerical modeling \citep{Sakurai2015}.  Since an accreting SMS stays in the supergiant protostar stage only with $\mdot \gtrsim 10^{-2}~\Msunyr$, the star may contract in a quiescent phase, so that its UV emissivity substantially rises. The stellar evolution calculations suggest that the contraction could occur if the duration of the quiescent phase is longer than $1000$ years. 
Multi-dimensional hydrodynamic simulations suggest that such a long break of the accretion is unlikely to occur, as long as the star is embedded in a massive self-gravitating disk \citep{Sakurai2016}.

%--------------------------------------------------------------------------------%

The above is all about the formation of SMSs postulated in the DC model. In the normal primordial star formation, the typical accretion rates $10^{-3} - 10^{-2}~\Msunyr$ are too low to bring an accreting star to the supergiant protostar stage. The protostar evolves through the normal KH contraction stage in this case, and the UV feedback is thought to become strong enough to halt the mass accretion. However, the variable mass accretion also commonly occurs in such a case, and may modify the above picture. \citet{Hosokawa2016} study the interplay between the variable accretion and resulting UV feedback, performing 3D radiation hydrodynamic simulations coupled with stellar evolution calculations. Consider the KH contraction stage when a bipolar HII region is growing (Fig.~\ref{ra_fig5}-a). Once the accretion rate exceeds $\sim 10^{-2}~\Msunyr$ with an accretion burst, a protostar rapidly inflates and its UV emissivity accordingly drops. The HII region disappears (b), and it is after a while in the quiescent phase that another HII region appears as the star contracts again (c). The cycle recurrently occurs as far as the disk accretes the gas from a surrounding envelope.
The UV feedback only operates intermittently, and it does not always easily halt the mass accretion.  The mass accretion continues even after the stellar mass exceeds $300~M_\odot$ in some cases.

%%%%%%%%%%%%%%%%%%%%%%%%%%%%%%%%%%%%%%%%%
\section{Summary and Future Prospects}
\label{ra_sec4}
%%%%%%%%%%%%%%%%%%%%%%%%%%%%%%%%%%%%%%%%%

We have described recent theoretical developments on the evolution of SMSs, highlighting effects of very rapid accretion at the rates of $\dot{M}_* \gtrsim 0.1~\Msunyr$.
Stellar evolution calculations predict that such a SMS has a very characteristic feature, i.e., large radius that monotonically increases as the stellar mass increases via the accretion. 
The evolution of the radius is well approximated as $R_* \propto M_*^{0.5}$, independent of the accretion rate if it exceeds a critical rate. The radius exceeds $7000~\rsun$ (or $30$~AU) after the star accretes the gas of $10^3~\Msun$. Interestingly, the stellar effective temperature is almost fixed at a constant value $T_{\rm eff} \simeq 5000$~K during this ``supergiant protostar'' stage. With the very cool atmosphere, the star only emits a small amount of ionizing photons, and the resulting UV feedback is too weak to disturb the accretion flow. 
The mass accretion onto the star thus continues even after the stellar mass reaches $\sim 10^4 - 10^5~\Msun$. 
While the mass accretion continues, the star finally collapses owing to
the exhaustion of nuclear fuel, or general relativistic instability.
Since the rapid accretion greatly affects the stellar interior structure, so it does also for its final fate.

%----------------------------------------------------------------------------%

The above picture has been established by numerically solving the 1D interior structure of an accreting SMS. However,  there still remains additional effects that need further investigations. 
One of them is potential mass-loss via strong stellar winds. If the star accretes the gas through a geometrically thin disk, most of the stellar surface is evacuated from it. 
We normally solve the structure of such a freely radiating atmosphere assuming the hydrostatic balance. However, relaxing the assumption allows the dynamical gas motion, so that very powerful stellar winds may be launched. If the resulting mass-loss rate exceeds the accretion rate, the stellar mass growth will be limited by this effect before reaching the supermassive regime. 
Recent theoretical studies suggest that the gas acceleration by radiation pressure and/or pulsational instability will excite some outflows, though the resulting mass-loss rates are much smaller than the accretion rates \citep{Inayoshi13, Nakauchi17}. 

%----------------------------------------------------------------------%

The angular momentum of the accreting gas may also play a critical role during the evolution of SMSs. It is still uncertain how much angular momentum should be carried into the star with the accreting gas, and how the resulting stellar evolution is modified by effects of the rotation. 
In general, the maximum rotation speed allowed for a star is set by the balance between the centrifugal force and gravity ($\Omega$-limit). However, a very massive star has the lower critical speed because a large contribution of the radiation pressure effectively reduces the gravity. 
If the star obtains the angular momentum carried by the accreting gas, the star easily reaches such a limit  \citep[known as $\Omega \Gamma$-limit, e.g.,][]{Meynet2000}. 
 The angular momentum of the star may be somehow extracted and transported outward through the disk while the star is accreting the gas~\citep{Paczynski91,Popham91}. 
Potential effects of this $\Omega \Gamma$-limit for the evolution of accreting SMSs have been controversial among recent studies \citep{Lee2016,Haemmerle17b,Takahashi17}, and are to be examined in detail in future studies. 

With this chapter, we conclude the discussion on the basic black hole formation scenarios and their progenitors. The next chapter~9 will summarize the statistical predictions on the first black holes. Their subsequent growth will be discussed in chapters~10 and 11. The final chapters will present both the current observational status as well as future prospects.

%\clearpage

%\bibliographystyle{ws-rv-van}
%\bibliography{biblio}

%\blankpage
%\printindex[aindx]                 % to print author index
%\printindex                         % to print subject index

%\end{document} 